\documentclass[12pt]{article}
\usepackage{amsmath, amsthm, amsfonts, bm}
\usepackage{amssymb}
\usepackage{mathtools}
\usepackage{float}

\usepackage{hyperref}
\hypersetup{colorlinks=true, citecolor=blue, linkcolor=blue, filecolor=blue}
\usepackage{graphicx,psfrag,epsf, float}
\usepackage{enumerate,titlesec, color}
\usepackage{natbib}
\usepackage{footnote}
\usepackage{url} % not crucial - just used below for the URL 
\usepackage{soul}
\usepackage{tikz}
\usepackage[ruled,linesnumbered]{algorithm2e}%[linesnumbered,lined,boxed,commentsnumbered, norelsize]
\usepackage[font=small,labelfont=bf]{caption}

\usepackage{lineno}

\usepackage{soul}

\usepackage{setspace}
\usepackage[textwidth=6.5in,bottom=1.8in,top=1in]{geometry}

\renewcommand\footnotemark{}

\usepackage{booktabs}

\usepackage{array}
\usepackage{multirow}
\definecolor{napiergreen}{rgb}{0.16, 0.5, 0.0}
\definecolor{myrtle}{rgb}{0.13, 0.26, 0.12}
\definecolor{dartmouthgreen}{rgb}{0.05, 0.5, 0.06}

\begin{document}
%%%%%%%%%%%%%%%%%%%%%%
%bold small letters
\def\bfa{\mathbf a}
\def\bfb{\mathbf b}
\def\bfc{\mathbf c}
\def\bfd{\mathbf d}
\def\bfe{\mathbf e}
\def\bff{\mathbf f}
\def\bfg{\mathbf g}
\def\bfh{\mathbf h}
\def\bfx{\mathbf x}
\def\bfy{\mathbf y}
\def\bfz{\mathbf z}
\def\bfs{\mathbf s}
\def\bfv{\mathbf v}
\def\bfr{\mathbf r}
\def\bv{\boldsymbol v}
\def\bu{\boldsymbol u}

\def\mv{\mathrm v}
\def\md{\mathrm d}
\def\mx{\mathrm x}

\def\wtbfc{\widetilde{\mathbf c}}
\def\wtbfgamma{\widetilde{\boldsymbol \gamma}}
\def\wtmx{\widetilde{\mathrm x}}
\def\wtbfbeta{\widetilde{\boldsymbol \beta}}
\def\wtmT{\widetilde{\mathrm T}}
\def\wtmH{\widetilde{\mathrm H}}
\def\wtmh{\widetilde{\mathrm h}}

\def\bw{\boldsymbol w}
%bold big letters
\def\bfA{\mathbf A}
\def\bfB{\mathbf B}
\def\bfC{\mathbf C}
\def\bfF{\mathbf F}
\def\bfS{\mathbf S}
\def\bfR{\mathbf R}
\def\bfI{\mathbf I}
\def\bfX{\mathbf X}
\def\bfE{\mathbf E}

\def\bfY{\mathbf Y}
\def\bfZ{\mathbf Z}
\def\bfD{\mathbf D}
\def\bfL{\mathbf L}
\def\bfM{\mathbf M}
\def\bfN{\mathbf N}
\def\bfw{\mathbf w}

\def\bfW{\mathbf W}
\def\bfU{\mathbf U}
\def\bfV{\mathbf V}
\def\bfu{\mathbf u}

%bold greek letters
\def\bfalpha{\boldsymbol \alpha}
\def\bfbeta{\boldsymbol \beta}
\def\bfkappa{\boldsymbol \kappa}
\def\bfepsilon{\boldsymbol \epsilon}
\def\bfdelta{\boldsymbol \delta}
\def\bfzeta{\boldsymbol \zeta}
\def\bfPhi{\boldsymbol \Phi}
\def\bfgamma{\boldsymbol \gamma}
\def\bfGamma{\boldsymbol \Gamma}
\def\bfTheta{\boldsymbol \Theta}
\def\bfmu{\boldsymbol\mu}
\def\bflambda{\boldsymbol\lambda}
\def\bfLambda{\boldsymbol\Lambda}
\def\bfmu{\boldsymbol\mu}
\def\bfomega{\boldsymbol\omega}
\def\bfphi{\boldsymbol\phi}
\def\bfpsi{\boldsymbol\psi}
\def\real{\mathbb R}

\def\bfvarepsilon{\boldsymbol\varepsilon}
\def\bfeta{\boldsymbol\eta}
\def\bfxi{\boldsymbol\xi}
\def\bfpi{\boldsymbol\pi}

\def\bfSigma{\boldsymbol\Sigma}
%bold number
\def\bfzero{\boldsymbol 0}
\def\bfone{\boldsymbol 1}
\def\bft{\mathbf t}

%%%%%%%%%%%%%%%%%%%%%%%%%%%%%%
% widetilde
\def\wtb{\widetilde b}
\def\wta{\widetilde a}
\def\wtc{\widetilde c}
\def\wtbeta{\widetilde\beta}
\def\wtgamma{\widetilde\gamma}
\def\wtmu{\widetilde \mu}
\def\wtbfmu{\widetilde {\boldsymbol\mu}}
\def\wtbfSigma{\widetilde {\boldsymbol\Sigma}}

%%%%%%%%%%%%%%%%%%%%%%%%%%%%%%
%math calligraphic
\def\cA{\mathcal A}
\def\cB{\mathcal B}
\def\cN{\mathcal N}
\def\cG{\mathcal G}
%%%%%%%%%%%%%%%%%%%%%%%%%%%%%%
%mathbb
\def\mbE{\mathbb E}
\def\mbZ{\mathbb Z}
\def\mbR{\mathbb R}
\def\mbGL{\mathbb {GL}}

%%%%%%%%%%%%%%%%%%%%%%%%%%%%%%
%mbox for distribution
\def\mLP{\mbox{LP}}
\def\mN{\mbox{N}}
\def\mIsing{\mbox{Ising}}
\def\mG{\mbox{G}}
\def\mU{\mbox{U}}
\def\mIG{\mbox{IG}}
\def\mT{\mathrm T}
\def\mH{\mathrm H}
\def\mP{\mathrm P}
\def\mh{\mathrm h}
\def\mBer{\mathrm{Ber}}
\def\mC{\mathrm C}
\def\cC{\mathcal C}
\def\cS{\mathcal S}
\def\cB{\mathcal B}
\def\cQ{\mathcal Q}
\def\cM{\mathcal M}
\def\cG{\mathcal G}
\def\mtr{\mbox{tr}}
\def\arrowPhi{\overrightarrow \Phi}
\def\linePhi{\overline \Phi}
%%%%%%%%%%%%%%%%%%%%%%%%%%%%%%
% define operations
\def\diag{\mbox{diag}}
\def\sign{\mbox{sign}}
\def\Pr{\mbox{Pr}}
\def\E{\mbox{E}}
\def\iid{\scriptsize \mbox{iid}}
\def\prox{\mbox{prox}}
\def\argmin{\mbox{argmin}}
\def\argmax{\mbox{argmax}}

%%%%%%%%%%%%%%%%%%%%%%%%%%%%%%
%
\newcommand{\obeta}[1]{\widetilde\beta_{r_{#1}}}
\newcommand{\isep}{\mathrel{{.}\,{.}}\nobreak}

	\bigskip
	\date{}

\title{Bayesian mixed model inference for genetic association under related samples with brain network phenotype}
\author{{Xinyuan Tian$^{1,\dagger}$}, {Yiting Wang$^{1,\dagger}$}, {Selena Wang$^1$}, {Yi Zhao$^2$}, {Yize Zhao$^{1}$}\bigskip\\
$^1$Department of Biostatistics, Yale University, New Haven, CT\\
$^2$Department of Biostatistics and Health Data Science, \\ Indiana University School of Medicine, Indianapolis, IN
 \footnote{$^\dagger$ Equally contributed}
% \footnote{$^*$ Correspondence should be directed to: yize.zhao@yale.edu}
}
%\email{yize.zhao@yale.edu}
	\maketitle

\def\spacingset#1{\renewcommand{\baselinestretch}%
	{#1}\small\normalsize} \spacingset{1}

 \vspace{-0.6cm}
\begin{abstract}
Genetic association studies for brain connectivity phenotypes have gained prominence due to advances in non-invasive imaging techniques and quantitative genetics. Brain connectivity traits, characterized by network configurations and unique biological structures, present distinct challenges compared to other quantitative phenotypes. Furthermore, the presence of sample relatedness in most imaging genetics studies limits the feasibility of adopting existing network-response modeling. In this paper, we fill this gap by proposing a Bayesian network-response mixed-effect model that considers a network-variate phenotype and incorporates population structures including pedigrees and unknown sample relatedness. To accommodate the inherent topological architecture associated with the genetic contributions to the phenotype, we model the effect components via a set of effect subnetworks  and impose an inter-network sparsity and intra-network shrinkage to dissect the phenotypic network configurations affected by the risk genetic variant. To facilitate uncertainty quantification of signaling components from both genotype and phenotype sides, we develop a Markov chain Monte Carlo (MCMC) algorithm for posterior inference. We evaluate the performance and robustness of our model through extensive simulations.  By further applying the method to study the genetic bases for brain structural connectivity using data from the Human Connectome Project with excessive family structures, we obtain plausible  and interpretable results. Beyond brain connectivity genetic studies, our proposed model also provides a general linear mixed-effect regression framework for network-variate outcomes.

\iffalse
 Non-invasive imaging techniques have made mapping the whole brain networks accessible. The current approaches often ignore the network structure of the brain connectivity by vectorizing the dependent connectivity edges and assuming independence among them, which results in substantial information loss. Moreover, few account for polygenetic effects, which can lead to estimation bias. In this article, we propose an integrative Bayesian framework to model the relationship between the single nucleotide polymorphism (SNP) and the brain structural connectivity and to identify genetic risk factors associated with brain connectome while taking into account people's family structure. To accommodate the biological architectures of brain connectivity constructed along white matter fiber tracts, we develop a symmetric matrix response regression of brain structural connectivity to characterize the association between genetic risk factors and brain connectome. We impose sparsity within graph and apply shrinkage between graph to identify informative networks and eliminate noise. Through simulations, we demonstrate the superiority of our method over existing approaches regarding the accuracy of estimating the effect elements and identifying the informative networks. We apply the proposed method to data from the Human Connectome Project (HCP) to obtain neurobiologically plausible insights with potential to inform future precision medicine.
\fi
\end{abstract}
\vspace{1cm}
\noindent%

\newpage
\spacingset{2}
\section{Introduction}
\label{s:intro}

Brain imaging genetics, aiming to uncover the genetic basis of brain structure and function, has provided an unprecedented opportunity to understand the molecular support for different neurobiological processes \citep{introd1_1}.   By leveraging imaging quantitative traits as endophenotypes that reflect underlying neurological etiologies, we gain a deeper understanding of the risk biomarkers implicated in both disease outcomes and normal trajectory of development and aging \citep{introd1_2,introd1_3}.

Brain connectivity, encoding the relations between distinct units or nodes within a nervous system, has played an essential role in disclosing the brain neuronal interactions and reflecting correspondence with behavior. Depending on the aspect of characterization, brain connectivity can be summarized by anatomical links capturing the white matter fiber tracts known as structural connectivity, or statistical dependence between functional time courses known as functional connectivity. Converging evidence indicates brain connectivity is heritable, and can offer distinct genetic underpinnings compared with other neuroimaging traits~\citep{zhao2021common,elliott2018genome}. This  underscores the significance of studying the  genetic contributions to connectivity patterns. From an analytical perspective, structural and functional connectivity can be viewed as an undirected graph with all the nodes over the brain as the vertex set and the corresponding connections as the edge set. By extracting single edges as univariate phenotypes, most of the current genome-wide association studies (GWAS) were performed separately on each brain connection~\citep{zhao2021common,jahanshad2013genome,elsheikh2020genome}. However, such analyses overlook the biological interdependence and graphical structure inherent in brain network topography, which can raise concerns regarding biological plausibility and interpretability, as our data application demonstrates.
In addition, this type of univariate approach finishes with a step of accounting for multiplicity, which significantly reduces the statistical power.

On the other hand, as the study of brain connectivity gains increasing interest, network-variate modeling has emerged as an advanced analytical framework capable of accommodating the underlying dependence and brain topological architectures. In contrast to marginal and univariate analyses, network-variate modeling directly handles the (weighted) adjacency matrix of connectivity, enabling an explicit characterization of the biological structure. Depending on the objectives of the study, the network-variate can serve three distinct roles. Firstly, it can be employed solely to describe neurobiological profiles of the brain using different types of graphical modeling techniques in light of topological assumptions~\citep{chen2023bayesian,wang2020locus,zhang2020mixed}. Secondly, when associated with a behavioral outcome, the network-variate can be treated as a predictor, involving specific matrix/tensor operations such as outer products~\citep{wang2021learning} to transform the predictive component into a linear term~\citep{zhao2022bayesian}. Finally, to investigate the impact of covariates or exposures on the variation of connectivity, the network variate can be treated as an outcome in a network-response regression. In this case, the coefficient parameters reveal a matrix or tensor format and can be further decomposed to elucidate the latent effect mechanisms~\citep{zhang2023generalized,zhao2023genetic,kong2019l2rm,hu2021nonparametric}. It is evident that the last category could shed light on genetic association analyses involving connectivity or network-variate phenotypes.

\iffalse
 The roles genetics play in explaining the individual differences in the brain structural connectome remains unclear, making it difficult to understand how brain structures form at different developmental stages and onsets of mental illness. Previous research often focuses on summarized network features or specific white matter fiber bundles \citep{intro4_11, intro4_12}. A few studies have linked brain functional connectivity to genetic variants and applied to a variety of subject matter fields \citep{intro4_21,intro4_22}. Although they are able to do what (improvement from second sentence), these methods do not account for the fact that genetic impact on the brain connectivity variation can be confounded with non-genetic effects, especially when using full-sib families produced by crossing two heterozygous parents. This confound may cause the estimated genetic variants to be inaccurate and biased \citep{kin1}. To address this issue, we propose an integrated Bayesian framework to model the association between Single-nucleotide polymorphism(SNP) and brain structural connectivity while accounting for the family relationships. By fitting the proposed model to data,  we demonstrate that the proposed model has a clear biological interpretation, enhancing its robustness and translational impact.
\fi
 
From a study design perspective, sample relatedness is highly prevalent and almost unavoidable in quantitative genetics studies. Such relatedness could be induced by recruitment from the same family or pedigree, or unknown or uncertain relationships including  distant levels of unknown common ancestry~\citep{eu2014comparison}.  Failure to account for potential sample structures within GWAS can lead to spurious results~\citep{helgason2005icelandic}, emphasizing the necessity for appropriate correction methods. One common approach to address sample structures is to include a random effect component to account for known or unknown relatedness. Building on linear mixed-effects models (LMMs), various  numerical implementation approaches proposed in recent years to characterize genetic associations accommodating population substructure and potential sample relatedness~\citep{kang2010variance,gemma}. However, most of these approaches are designed for univariate phenotypes or vector-variate multivariate phenotypes, and there is currently no existing framework that adequately considers or readily applies to network-variate phenotypes.

To address the above limitations, we propose a \textbf{B}ayesian \textbf{N}etwork-phenotype \textbf{M}ixed \textbf{E}ffect model (BNME)  to perform genetic association analyses with brain connectivity phenotype. Within this unified modeling framework, we simultaneously characterize genetic contributions and identify affected phenotypic network components, while quantifying their uncertainty. To leverage the biological knowledge that brain connectivity operates via subnetwork configurations, our approach assumes that risk genetic variants influence network alternations by acting upon specific subnetwork/subnetwork units that are to be uncovered. By imposing shrinkage and sparsity priors on the effect parameters, we can map out the genetically targeted brain subnetworks that play a critical role to guide future intervention strategies. In contrast to recent works on network-response genetic association analyses including \cite{zhao2023genetic} and \cite{kong2019l2rm}, our proposed method incorporates pedigree information and accounts for sample structures, ensuring the reliability and validity of the findings. In our data application, we apply the BNME model to study the genetic bases of brain structural connectivity in the Human Connectome Project (HCP), accommodating its extensive family structures. 
Lastly, despite the proposed model being motivated by brain connectivity genetic studies, it can be readily extended to perform general network- or matrix-response mixed effects modeling. To the best of our knowledge, this work is among the very first to develop such a modeling framework, which directly fulfills an urgent need to capture multi-source of random variability for a growing collection of network data in biomedical studies.

The remainder of the article is organized as follows. In Section~\ref{sec: method}, we describe the proposed LMM with a network response (Section~\ref{sec:LMM}), the prior specifications (Section~\ref{sec:prior}), the posterior inference procedure (Section~\ref{sec:posterior}), and the non-genetic effect adjustment (Section~\ref{sec:cov}). We conduct simulation studies to evaluate the proposed model compared with existing alternatives in Section~\ref{sec:simu}, followed by an application to HCP brain connectivity genetics data in~Section \ref{sec:data}. In the end, we conclude the paper with a discussion in Section~\ref{s:discuss}.

\section{Method}\label{sec: method}

\subsection{Linear mixed-effect model with a network phenotype}\label{sec:LMM}
We first describe the problem setting in the context of GWAS with genetic correlation, though the model formulation represents a general network-response mixed-effect model that can be extended to other applications. Assume the study includes $N$ subjects with known pedigree structure or unknown relationship.  For subject $i~ (i=1,\dots,N)$, let $z_i$ denote the genotype of interest which is encoded as 0, 1 or 2 according to the number of copies for the tested allele,  $\bfx_i\in\mathbb{R}^{P\times 1}$ represents a set of covariates, and $\bfA_i\in\mathbb{R}^{V\times V}$ denotes the network phenotype summarized by a graphical matrix. With $\bfA_i$ stacked across all the subjects, we have the network phenotype array $\mathcal{A}\in\mathbb{R}^{V\times V\times N}$. Specifically in the application of brain network studies, with images processed under a common brain atlas with $V$ nodes, both structural and functional brain connectivity can be viewed as an indirect graph across vertex set $\{1, \dots, V\}$. Thus, $\bfA_i$ becomes a symmetric matrix to summarize brain connectivity for each subject with diagonal elements to be zero, and its $(k,l)$th entry $a_{ikl}$, $0< k \neq l \leq V$ represents the connection between nodes $k$ and $l$ characterizing either the white matter fiber tracts (structural connectivity) or statistical dependence of functional time course (functional connectivity). We adopt continuous metrics to measure structural and functional connections. After normalizing the genetic variant and each phenotypic connection, we propose the following genetic association model for the indirect network response,
\begin{align}
\bfA_i=\Theta z_i-\mbox{Hol}[\Theta z_i]+(\bfB_i-\bfB^T_i)+(\bfE_i-\bfE^T_i). \label{eq:matrix}
\end{align}
Here,  $\Theta\in\mathbb{R}^{V\times V}$ is the symmetric coefficient matrix to capture the genetic effect on the network phenotype, $\mbox{Hol}[ \cdot ]$ is the operation to hollow out the diagonal elements to form a diagonal matrix, $\bfB_i\in\mathbb{R}^{V\times V}$ is the symmetric random polygenic effect matrix, and $\bfE_i\in\mathbb{R}^{V\times V}$ is the symmetric random error matrix characterizing the environmental effects. To demonstrate the main idea, we include only genetic fixed effect at this moment, and we will extend the model to include non-genetic covariates afterwards. Model~\eqref{eq:matrix} can be viewed as an extension of the traditional linear mixed effect model for genetic association with univariate or multivariate phenotypes accommodating sample relatedness. In addition to  a matrix-variate phenotype, we design both mean and variance components to maintain their original functions while satisfying the symmetric and hollow structure of the indirect network as shown in the right-hand side of~\eqref{eq:matrix}. Specifically, for the genetic and environmental effect matrix, by stacking each of them across all the subjects, we have the random effect tensor $\mathcal{B}\in\mathbb{R}^{N\times V\times V}$ and residual error tensor $\mathcal{E}\in\mathbb{R}^{N\times V\times V}$ with
\begin{align*}
\mbox{vec}(\mathcal{B})&\sim \mbox{N}\left\{\begin{pmatrix}
   0\\ 
 \vdots \\ 
0
 \end{pmatrix},   \mbox{Diag}\begin{pmatrix}
   \sigma^{(a)}_{11}\\ 
 \vdots \\ 
\sigma^{(a)}_{VV}
 \end{pmatrix}\otimes\Lambda\right\}\label{eq:matrix1}, \quad
\mbox{vec}(\mathcal{E})\sim \mbox{N}\left\{\begin{pmatrix}
   0\\ 
 \vdots \\ 
0
 \end{pmatrix},  \mbox{Diag}\begin{pmatrix}
   \sigma^{(e)}_{11}/2  \\ 
  \vdots  \\ 
\sigma^{(e)}_{VV}/2 
 \end{pmatrix}\otimes\mbox{I}_N\right\}, 
\end{align*}
where $\mbox{Diag}(\cdot)$ constructs the diagonal matrix formed by the inside vector, $\mbox{I}_N\in\mathbb{R}^{N\times N}$ is the identify matrix, and $\Lambda\in\mathbb{R}^{N\times N}$ is the kinship matrix estimated by pedigree information for known family structures or genotypic relationship for unknown relatedness \citep{eu2014comparison}. By proposing so, we can show the phenotypic variance of each connection $\mbox{Var}(\bfa_{kl})=2\sigma^{(a)}_{kl}\Lambda+\sigma^{(e)}_{kl}\bfI_N$, $\bfa_{kl}=(a_{1kl},\dots, a_{Nkl})^T, k \neq l$, consistent with the existing literature \citep{kang2010variance}.

Given the size of commonly used brain atlas can be large with $V$ in the range of 200 to 1000, directly performing estimation on model \eqref{eq:matrix} is not ideal under a high-dimensional parameter space. More importantly, considering the primary interest to investigate the genetic association with brain network architectures, the topological structure can not be plausibly reflected by ignoring the dependence within the genetic coefficient matrix. To address so, we adopt the following Tucker decomposing under a symmetry constrain for the coefficient matrix
\begin{equation}
\Theta = \sum_{h=1}^H \eta_h\boldsymbol{\theta}_h\circ\boldsymbol{\theta}_h,\label{eq:matrix1}
\end{equation}
where $\circ$ represents the outer product, and $\boldsymbol{\theta}_h=(\theta_{h1},\dots,\theta_{hV})^T, h=1,\dots, H$ are column coefficient vectors. From a neurobiological perspective, each $\boldsymbol{\theta}_h\circ\boldsymbol{\theta}_h$ describes an effect network component adjusted by a weight parameter $\eta_h$.  Combining models~\eqref{eq:matrix} and~\eqref{eq:matrix1}, we allow the genetic variant delivers its impact on the phenotype via a series of signaling network architectures.

\subsection{Prior specifications}\label{sec:prior}

We consider a fully Bayesian paradigm to estimate and perform inference for the proposed network-response LMM. For the fixed genetic effect component, we anticipate the genetic impact is sparse across the brain as shown by the existing empirical studies~\citep{zhao2021common}. Therefore, we assign the following combination of point mass mixture prior and shrinkage prior
\begin{equation}\label{eq:prior1}
  \eta_{h}\sim (1-\tau_{h})\delta_0+\tau_{h}\mbox{N}(0,\omega); \quad \theta_{hv} \sim  \mathcal{L}(\nu), \quad h=1,\dots,H; v=1,\dots,V.
\end{equation}
Here, $\tau_{h}$ is the latent selection indicator to determine whether a network component is significantly impacted by the genotype as a whole. When $\tau_h=1$, the weight parameter $\eta_{h}$ is generated from a noninformative Normal prior with a large variance parameter $\omega$; otherwise, we assign  $\eta_{h}$ to a point mass at zero denoted by $\delta_0$ to remove the whole component from the model. In real practice, with the number of effect component $H$ unknown, such a specification of sparsity could efficiently assist the determination of the number of associated phenotypic network configurations during the learning process. As shown in our numerical studies, by imposing a conservative value to $H$, our model can correctly uncover the signaling network phenotypes. To specify priors for latent indicators $\tau_{h}$, one can either impose a non-informative Bernoulli distribution for each of the elements, or resort to a more informative prior by incorporating additional biological structure~\citep{li2010bayesian}. For the coefficients, we assign a Laplace prior $\mathcal{L}(\nu)$ with a scale parameter $\nu$ to shrink the noise effect to a close to zero value. To further facilitate a straightforward posterior computation, following \cite{park2008bayesian}, we represent each Laplace prior by a scale mixture of normals for each $h=1,\dots, H$,
\begin{align}\label{eq:prior2}
\boldsymbol{\theta}_h \sim \mbox{N}(\bfzero, \mathcal{D}_{h}); \quad \mathcal{D}_{h}=\mbox{Diag}(\sigma_{h1},\dots,\sigma_{hV}); \quad
    \sigma_{hv} &\sim \frac{\nu^2}{2}\exp(-\frac{\nu^2\sigma_{hv}}{2})d\sigma_{hv}, v=1,\dots,V.
\end{align}
Combining priors~\eqref{eq:prior1} and~\eqref{eq:prior2}, we characterize the phenotypic signals in a hierarchical way with a group-level sparsity to induce the selection of a phenotypic network as a whole and a within-group shrinkage to identify the actual signaling subnetwork structure within selected ones. In contrast to existing sparse group selection or shrinkage models that primarily focus on group structural covariates \citep{simon2013sparse,chang2018scalable},our research emphasizes the network-variate outcome, which captures the associations between covariates and latent topological hierarchies. Additionally, we opt for shrinkage priors for individual coefficients instead of point mass mixture priors, driven by computational considerations that result in lower computational costs for shrinkage priors. However, it is important to note that the Laplace prior can be readily replaced with spike-and-slab types of priors or other graphical priors~\citep{chang2018scalable,stingo2011incorporating} to impose  sharp sparsity or incorporate spatial information.  Finally, 
we assign non-informative inverse gamma (IG) prior for variances $\sigma^{(a)}_{kl}, \sigma^{(e)}_{kl}$ with shape and scale parameters $\alpha$ and $\beta$, respectively, and a relatively large value for the rest of the variance parameters during implementation.  For the tuning parameters including the number of informative subnetworks $H$ and scale parameter $\nu$, we consider a grid search of them and choose the optimal values using the Bayesian information criterion (BIC). Our numerical experience suggests that this strategy is effective in practical applications.

\subsection{Posterior likelihood and inference for BNME}\label{sec:posterior}

To perform posterior inference for the proposed BNME model, we first develop the posterior likelihood for the collect of unknown parameters  denoted as $\bfzeta=\bigg[\big\{\boldsymbol{\theta}_h, \eta_h,\tau_h, \big(\sigma_{hv}\big)_{v=1}^V\big\}_{h=1}^H, \\ \big(\sigma^{(a)}_{vv'}\big)_{v,v'=1}^V, \big(\sigma^{(e)}_{vv'}\big)_{v,v'=1}^V, \nu\bigg]$.  Based on the observed data $\mathcal{O}=(\bfA_i, z_i, \Lambda; i=1,\dots, N)$, the joint posterior distribution follows 
\begin{equation}
\begin{aligned}
\pi(\bfzeta\mid \mathcal{O})\propto& \prod_{i}\pi\left(\bfA_i , z_i, \Lambda\mid \{\boldsymbol{\theta}_h, \eta_h,\tau_h\}_{h=1}^H, \big(\sigma^{(a)}_{vv'}\big)_{v,v'=1}^V, \big(\sigma^{(e)}_{vv'}\big)_{v,v'=1}^V \right)\\
&\times\prod_{v} \bigg\{\prod_{h}\pi(\boldsymbol{\theta}_{hv}\mid \sigma_{hv},\nu)\pi( \sigma_{hv})\prod_{v'}\pi(\sigma_{vv'}^{(a)})\pi(\sigma_{vv'}^{(e)})\bigg\}\prod_{h} \bigg\{\pi(\tau_h)\pi(\eta_h|\tau_h)\bigg\},\label{eq:post}
\end{aligned}
\end{equation}
which combines the conditional observed data likelihood with prior distributions. Given uncertainty quantification is an essential component for genetic association analyses, instead of pursuing point estimates via optimization algorithms,  we develop a Markov chain Monte Carlo (MCMC) sampling algorithm for posterior inference based on a combination of Gibbs samplers and Metropolis-Hastings (MH) updates.  Under random initializations, we cycle through the following steps:
\begin{itemize}
\item For $h=1\dots H$, $k=1\dots V$, $l= 1\dots V$, denote the $(k,l)$th entry  of matrix $\{\bfA_i-\sum_{h'\neq h} \eta_{h'}\boldsymbol{\theta}_{h'}\circ\boldsymbol{\theta}_{h'}z_i\}$ as $\widetilde{a}_{iklh}$, and define $\widetilde{\boldsymbol{a}}_{klh}=(\widetilde{a}_{1klh},\dots,\widetilde{a}_{Nklh})^T$, $\boldsymbol{z}=(z_1,\dots,z_N)^T$. Sample $\theta_{hk}$ from $\mbox{N}(\mu_{{\theta}_{hk}},\sigma_{\theta_{hk}})$ with
$\sigma_{{\theta}_{hk}}=\big(\sum_{l\neq k}\eta^2_h\theta^2_{hl}\boldsymbol{z}^T\left(\sigma^{(a)}_{kl}\Lambda+\sigma^{(e)}_{kl}I\right)^{-1}\boldsymbol{z}+\sigma^{-1}_{hk}\big)^{-1}$ and $\mu_{\theta_{hk}}=\sum_{l\neq k}\widetilde{\boldsymbol{a}}_{klh}^T\Big(\sigma^{(a)}_{kl}\Lambda+\sigma^{(e)}_{kl}\bfI\Big)^{-1}\boldsymbol{z}\eta_h\theta_{hl}\sigma_{\theta_{hk}}$

 \item For $h=1\dots H$, $k=1\dots V$, sample $\sigma_{hk}^{-1}$ from an Inverse Normal distribution $\mbox{IN}(\frac{\nu}{\vert\theta_{hk}\vert},\nu^2)$.

\item For $h=1\dots H$, when $\tau_h=0$, set $\eta_h$ to be zero. Otherwise, denote the $(k,l)$th entry  of matrix $\boldsymbol{\theta}_{h}\circ\boldsymbol{\theta}_{h} z_i$ as $q_{iklh}$, and $\boldsymbol{q_{klh}}=(q_{1klh},\dots,q_{Nklh})^T$. Update between subnetwork coefficient  $\eta_h$ from their corresponding posterior Normal distribution $\mbox{N}(\mu_{\eta_h},\sigma_{\eta_h})$ with
   $\sigma_{\eta_h}=\left(\sum_{k<l}\boldsymbol{q_{klh}}^T(\sigma^{(a)}_{kl}\Lambda+\sigma^{(e)}_{kl}I)^{-1}\boldsymbol{q_{klh}}+\omega^{-1}\right)^{-1}$, and $\mu_{\eta_h}=\sum_{k<l}\widetilde{\boldsymbol{a}}_{klh}^T(\sigma^{(a)}_{kl}\Lambda+\sigma^{(e)}_{kl}I)^{-1}\boldsymbol{q_{klh}}\Sigma_{\eta_h}$.

\item For $h=1\dots H$, define $l_0:=\frac{C}{\omega}\exp(-\frac{1}{2}(\frac{C\eta_h}{\omega})^2)$ and $l_1:=\frac{1}{\omega}\exp(-\frac{1}{2}(\frac{\eta_h}{\omega})^2)$ with  $C$ a large constant. We then update the selection indicators $\tau_h$ following the posterior Bernoulli distributions Bern($\frac{l_1}{l_0+l_1}$).

\item For $v=1\dots V$, $v'=1\dots V$, update $\sigma_{vv'}^{(a)}$ by sampling a proposed value $\sigma_{vv'}^{(a)p}$ from a random walk proposal distribution $\mbox{N}(\sigma_{vv'}^{(a)},\rho_1^2)$, and setting $\sigma_{vv'}^{(a)}=\sigma_{vv'}^{(a)p}$ with probability $\min\{1, R_1\}I\{\sigma_{vv'}^{(a)p}>0\}$, where
    $R_1 = \frac{\pi( \sigma_{vv'}^{(a)p}\mid \mathcal{O}, \{\eta_h,\tau_h, \boldsymbol{\theta}_{h}\}_{h=1}^H,\sigma^{(a)}_{k\neq v,l\neq v'}, \sigma^{(e)})}{\pi(\sigma_{vv'}^{(a)} \mid \mathcal{O}, \{\eta_h,\tau_h, \boldsymbol{\theta}_{h}\}_{h=1}^H,\sigma^{(a)}_{k\neq v,l\neq v'}, \sigma^{(e)})}$,
    with $\pi(\sigma_{vv'}^{(a)}\mid \mathcal{O}, \{\eta_h,\tau_h, \boldsymbol{\theta}_{h}\}_{h=1}^H,\sigma^{(a)}_{k\neq v,l\neq v'}, \sigma^{(e)})\propto\mathcal{L}(\mathcal{O}\mid \bfzeta)\{\sigma_{vv'}^{(a)}\}^ {-\alpha-1}\exp(-\frac{\beta}{\sigma_{vv'}^{(a)}})$ the full conditional.

\item For $v=1\dots V$, $v'=1\dots V$, update $\sigma_{vv'}^{(e)}$ by sampling a proposed value $\sigma_{vv}^{(e)p}$ from a random walk proposal distribution $\mbox{N}(\sigma_{vv'}^{(e)},\rho_2^2)$ and setting $\sigma_{vv'}^{(e)}=\sigma_{vv'}^{(e)p}$ with probability $\min\{1, R_2\}I\{\sigma_{vv'}^{(e)p}>0\}$, where
    $R_2 = \frac{\pi( \sigma_{vv'}^{(e)p}\mid \mathcal{O}, \{\eta_h,\tau_h, \boldsymbol{\theta}_{h}\}_{h=1}^H,\sigma^{(a)}, \sigma^{(e)}_{k\neq v,l\neq v'})}{\pi(\sigma_{vv'}^{(e)} \mid \mathcal{O}, \{\eta_h,\tau_h, \boldsymbol{\theta}_{h}\}_{h=1}^H,\sigma^{(a)}, \sigma^{(e)}_{k\neq v,l\neq v'})}$,
    with $\pi(\sigma_{vv'}^{(e)}\mid \mathcal{O}, \{\eta_h,\tau_h,\boldsymbol{\theta}_{h}\}_{h=1}^H,\sigma^{(a)}, \sigma^{(e)}_{k\neq v,l\neq v'})\propto\mathcal{L}(\mathcal{O}\mid \bfzeta)\{\sigma_{vv'}^{(e)}\}^ {-\alpha-1}\exp(-\frac{\beta}{\sigma_{vv'}^{(e)}})$ the full conditional.
\end{itemize}
Based on the posterior samples, the convergence of the algorithm is examined by trace plots and GR method~\citep{gelman1992inference}. To characterize the genetic impact and dissect the associated signaling brain network configurations,  we first determined the overall phenotypic subnetworks linked with the genetic variant based on a 0.5 cutoff of the  posterior mean for each $\tau_h$. This cutoff is adopted in light of the  median
probability model~\citep{hastie2004}. Under a conservative $H$, most of the risk genetic variants are associated with less than $H$ brain connectivity subnetworks. When none of the elements in $\{\tau_h\}_{h=1}^H$ surpasses the cutoff, the genetic variant does not provide a significant impact on any component of the network phenotype. For the selected subnetworks with $\tau_h$ larger than the cutoff, the genetic effect over network structures is captured by the posterior mean of $\theta_h$. Despite that a Laplace prior does not impose strict sparsity, we can determine the specific brain network configurations that are most relevant to the genetic impact by extracting the elements from $\theta_h$ with a credible interval excluding zero. Eventually, our model could provide estimation and inference for the risk genetic factors and their most influencing phenotypic topological elements.

\subsection{Covariates adjustment}\label{sec:cov}
In genetic association studies, it is a standard practice to adjust for non-genetic covariates, such as demographics and genetic principle components. In this paper, extending an existing projection approach for multivariate outcomes~\citep{fixedeffect,zhao2022bayesian}, we account for the effect of covariates in addition to model~\eqref{eq:matrix}.  Denote covariate matrix as $\boldsymbol{X}=(\boldsymbol{x}_1,\dots,\boldsymbol{x}_N)^T\in \mathbb{R}^{N\times P}$, and we define a projection matrix  
$\boldsymbol{W}=\bfI_{N\times N}-\boldsymbol{X}(\boldsymbol{X}^T\boldsymbol{X})^{-1}\boldsymbol{X}^T$. Clearly, $\boldsymbol{W}$ is symmetric and idempotent matrix with a rank of $(N-P)$, and this further indicates that $\boldsymbol{W}$ can be decomposed as $\boldsymbol{W}=\boldsymbol{U}^T\boldsymbol{U}$, where matrix $\boldsymbol{U}\in \mathbb{R}^{(N-P)\times N}$ and satisfies $\boldsymbol{U}\boldsymbol{U}^T=\bfI_{N-P}$ and $\boldsymbol{U}\boldsymbol{X}=\bfzero$. Through matrix $\boldsymbol{U}$, the data can be projected from the $N$ dimensional space onto an $(N-P)$ dimensional subspace. This faciliates an efficient way to remove the nuisance covariate effects by projecting both sides in the original model via 
\begin{align}\label{eq:model2}
\boldsymbol{U}\mathcal{A}=\bfTheta\boldsymbol{U}z+\boldsymbol{U}\mathcal{B}+\boldsymbol{U}\mathcal{E}.  
\end{align}
Model~\eqref{eq:model2} indicates that by replacing the connectivity array $\mathcal{A}$ with $\widetilde{\mathcal{A}}=\boldsymbol{U}\mathcal{A}$, genotype $z$ with $\widetilde{z}=\boldsymbol{U}z$ and the kinship matrix $\Lambda$ with $\widetilde{\Lambda}=\boldsymbol{U}\Lambda \boldsymbol{U}^T$,  the joint posterior distribution will follow the same structure as~\eqref{eq:post}. Hence, all sampling procedures can be adopted accordingly.

\section{Simulation Studies} \label{sec:simu}

\begin{figure}[H]
\centering
  \includegraphics[ width=0.47\textwidth]{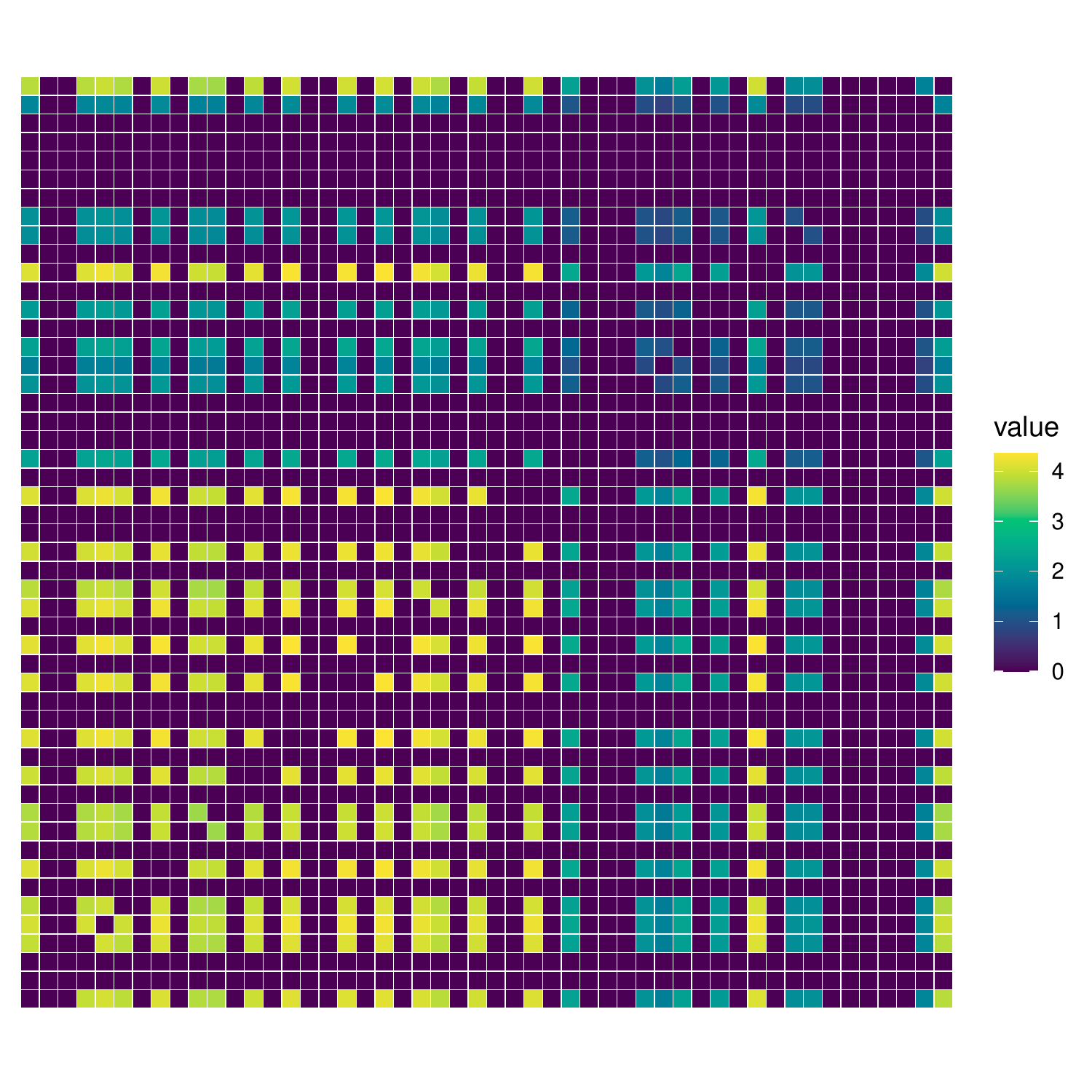}  
  \includegraphics[ width=0.47\textwidth]{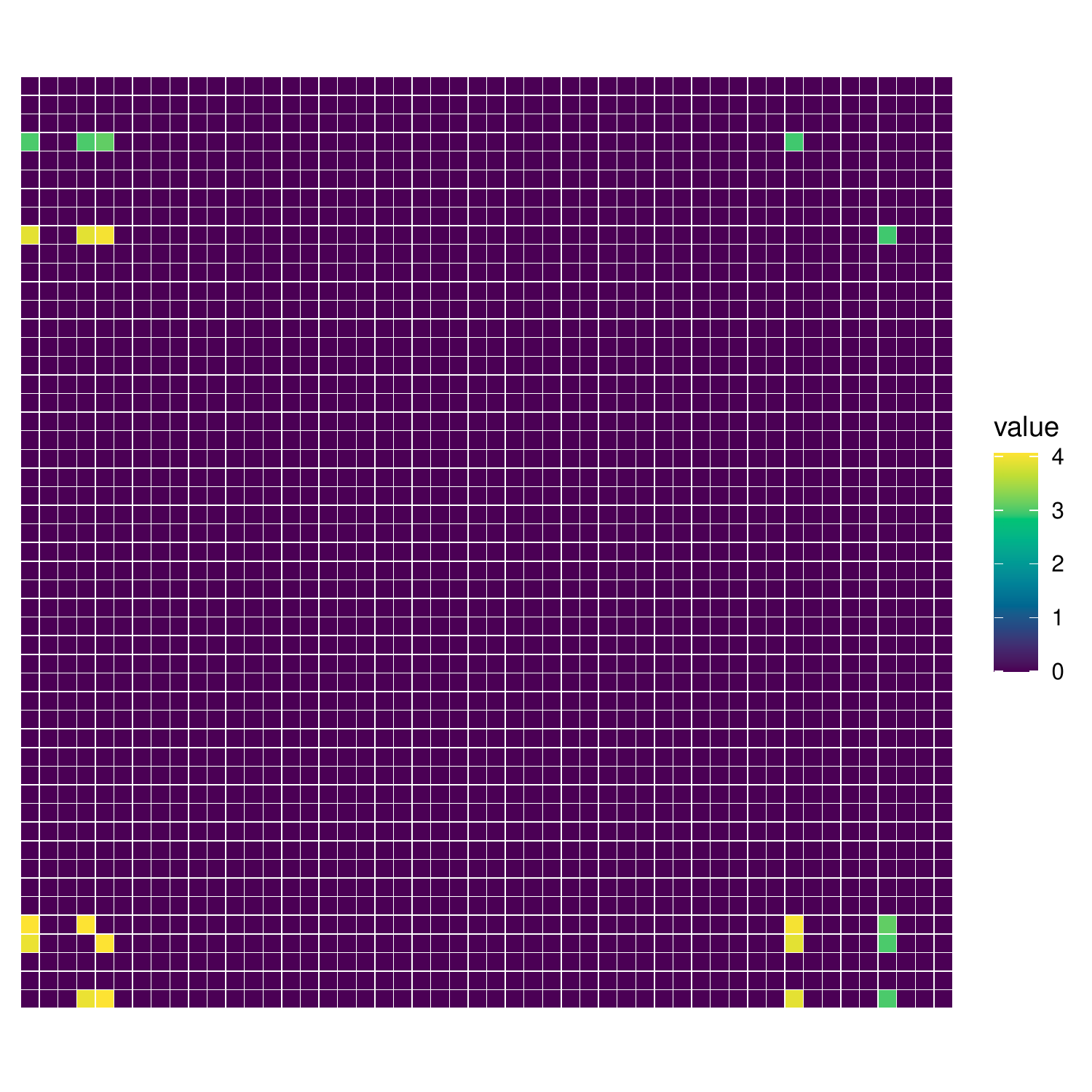}
 \caption{Assembled signal patterns upon the whole phenotyptic network at the sparsity levels of 50\% (left panel) and 90\% (right panel) under the second scenario with three subnetworks.}\label{fig_sig}
\label{signal}
\end{figure}

We carry out simulation studies to evaluate the proposed BNME to uncover genetic signals and the associated phenotypic network configurations under related samples. To mimic the data dimension in our data application, we assign sample sizes $N=100$ and $500$ with brain connectivity generated under a brain atlas with $V=50$. We consider two scenarios on the phenotypic network configurations that are highly impacted by the genetic factor. In the first scenario, we generate a single phenotypic subnetwork that is linked with the genetic variant, and we set $\eta_1=1$. In the second scenario, we create a more challenging setting by generating three subnetworks  with the associated weight parameter $\eta_h$ equals 0.7, 0.3 and 0, respectively. The third subnetwork is not linked to the genetic variant, allowing us to evaluate the performance of our model in detecting the true number of signaling phenotypic components. For both scenarios, we consider a range of sparsity levels for each  $\boldsymbol{\theta}_h$ by imposing 50\%, 90\% and 100\% of the elements within the vector to be zero to define the genetically associated clique subnetworks. As shown in Figure~\ref{fig_sig},  we provide the signal patterns upon the whole network phenotype under  50\% and 90\% sparsity levels for the second scenario assembled across subnetworks. Of note, when sparsity level is 100\%,  the genotype does not impact any of the phenotypic structures, facilitating a test on a noise genetic variant. For the variance components, we first generate a kinship matrix $\Lambda$ with diagonal entries to be 1 and off-diagonal entries ranging from (0,1) to be consistent with real practice. Then we set each of the polygenic effect variance $\sigma_{vv'}^{(a)}$ to be 1.5 and the environmental effect variance $\sigma_{vv'}^{(e)}$ to be 1. Finally, for the fix effects, we sample the genotype for each subject from $\left \{ 0, 1, 2  \right \}$, and add three different types of non-genetic covariates including one generated from a Bernoulli distribution $\mbox{Bern}(0.5)$, one from a Uniform distribution $\mathcal{U}(-0.5,0.5)$, and one from a  Normal distribution $\mathcal{N}(-0.5,0.5)$. Each of the fixed effect coefficients are generated from $\mathcal{N}(0.3, 0.5)$ and fixed for all the settings. Overall, we consider 12 settings with different sample sizes and phenotypic signal patterns, and we generated 200 Monte Carlo datasets for each setting.

To implement the proposed BNME model, we set $\alpha=\beta=0.01$ for a noninformative support of IG priors. To assess the robustness of the model, we directly set $H=3$ which is larger than the actual number of the associated phenotypic subnetworks for both scenarios, and we determine $\nu$ by a grid search from (0.5, 0.8, 1) based on BIC. The MCMC algorithm is performed for $5,000$ iterations after $2,000$ burn-in, and both trace plots and GR value indicate a convergence. For the competing methods, given there is no existing regression approach that can accommodate a network outcome with mixed effects, we extract unique edges from the  phenotype matrix. With each of the upper diagonal elements of $A_i$ as a phenotypic trait, we implement a linear mixed-effect model (LMM) using the {\tt lme4} package in {\tt R}, linear mixed-effects kinship model (LMEKIN) using the {\tt coxme} package in {\tt R} and one of the most popular GWAS pipelines for related samples and univariate phenotype Genome-wide Efficient Mixed Model Association~\cite[GEMMA,][]{gemma}.
To evaluate both estimation and feature selection, 
we consider the following performance metrics: (a) root mean predicted square error (RMSE) of $\bfA_i$, (b) sensitivity ($\text{Sen}_e$) and specificity ($\text{Spe}_e$) for distinguishing signaling phenotypic elements captured by the nonzero elements in $\Theta$, and (c) specificity ($\text{Spe}_g$) for identifying noise genetic variant when sparsity level is  $100\%$. All the simulation results are summarized in Table~\ref{table:simnew3}.

% I will revise the following paragraph later
Based on the results, we conclude that our proposed BNME model demonstrates excellent performance in uncovering genetic effects, identifying associated phenotypic network configurations, and distinguishing noise genetic variants. Specifically, the BNME exhibits a significantly smaller RMSE compared to alternative methods indicating higher estimation accuracy. Our method also achieves over 90\% phenotypic sensitivity and specificity across all the simulation settings, and genotypic specificity when the sparsity level is 100\%, indicating its ability to uncover the associated phenotypic networks for the risk genotype and distinguish the noise genetic variant. When comparing different settings, we consistently observe improvements in performance metrics for all methods as the sample size increases. As anticipated, a higher sparsity level aids in signal identification for all the methods. Notably, when sparsity reaches 100\% with no associated phenotypic connections, given that our method allows to exclude the noise phenotypic component entirely, it successfully detects this situation as evidenced by a  close to one $\text{Sen}_g$. Moveover, as more phenotypic subnetworks are impacted, including a noise subnetwork, we observe a notable decrease in the accuracy of phenotypic feature selection for all competing methods. However, our method maintains its superior performance, indicating its robustness and ability to uncover the true signaling phenotypic subnetworks even under a misspecified subnetwork number $H$. Finally, in the comparison among competing methods, both LMEKIN and GEMMA demonstrate similar performance, surpassing the traditional LMM. Their performance in the presence of  a noise genotype suggests a high risk of false positives when considering GWAS under a network phenotype.

\iffalse

When the sparsity level comes to 100\%, the sensitivity does not exist, as a result of no truely active subnetwork, which means the genotype $z$ is not associated with phenotype $A$. The results of 100\% sparsity scenarios show less than 0.1 in RMSE and approximately 100\% in specificity with extremely low standard deviation, which demonstrate that our model can identify the unrelated genotypes more precisely and accurately.
BNME and GEMMA consistently outperforms other shrinkage models in identifying the phenotypic network. This provides the evidence for the benefit of incorporating relatedness. The improvement of BNME against GEMMA becomes evident when we consider more subnetworks with a more sparse settings. One potential reason is that BNME can better handle sparse genetic impacts through the algorithm. Secondly, as we hypothesize an over-complicated three-subnetwork structure with two truely associated subnetwork, BNME can successfully detect the noise subnetwork, which is critical in constructing the accurate genetic associated network structure. 
In general, BNME shows a strong performance in identifying siginificant associated genotype and capturing the accurate phenotypic network information.
\fi
\begin{table}[!ht] \caption []{Simulation results for all the methods under different settings range from sample sizes ($N$), sparsity levels and phenotypic subnetworks (\# Sub). The results are summarized over 200 MC datasets and the standard deviations are included in the parenthesis. }\label{table:simnew3}
 \vspace{0cm}
\centering
\resizebox{\textwidth}{!}{
   \begin{tabular}{ccccccccccc}
\toprule 
&&&\multicolumn{4}{c}{N=100} & \multicolumn{4}{c}{N=500} \\
\cmidrule(lr){4-7} \cmidrule(lr){8-11}
{\# Sub} &  Sparsity  &  Model   &RMSE  &$\text{Spe}_e$ &$\text{Sen}_e$ &$\text{Spe}_g$ &RMSE&$\text{Spe}_e$ &$\text{Sen}_e$ &$\text{Spe}_g$\\
\cmidrule(lr){1-11}

&&BNME&0.13 (0.05)&0.96 (0.04)&1.00 (0.00)&- &0.04 (0.02)&0.97 (0.03)&1.00 (0.00)&-  \\
\cmidrule(lr){4-11}
&\textbf{\multirow{2}{*}{50\%}}&LMM&0.71 (0.22)&0.94 (0.12)&0.86 (0.10)&- &0.32 (0.06)&0.95 (0.05)&0.98 (0.01) &-\\
\cmidrule(lr){4-11}
&&LMEKIN&0.25 (0.10) &0.94 (0.14) &0.93 (0.05) &- &0.24 (0.07) &0.95 (0.05) &0.98 (0.03) &-  \\
\cmidrule(lr){4-11}
&&GEMMA&0.25 (0.13)&0.94 (0.14)&0.93 (0.03) &- &0.20 (0.04) &0.95 (0.05) &0.99 (0.00) &- \\

\cmidrule(lr){2-11}
&&BNME&0.53 (0.14)&0.99 (0.01)&1.00 (0.00)&- &0.34 (0.30)&0.99 (0.01)&1.00 (0.00) &-\\
\cmidrule(lr){4-11}
\textbf{\multirow{2}{*}{1}}&\textbf{\multirow{2}{*}{90\%}}&LMM&0.58 (0.02)&0.94 (0.01)&0.99 (0.02)&- &0.35 (0.26)&0.95 (0.04)&1.00 (0.00) &-  \\
\cmidrule(lr){4-11}
&&LMEKIN&0.26 (0.07) &0.93 (0.03) &0.99 (0.01) &- &0.28 (0.06) &0.95 (0.05) & 0.99 (0.00) &-  \\
\cmidrule(lr){4-11}
&&GEMMA&0.23 (0.09) &0.95 (0.06)&0.99 (0.01) &- &0.22 (0.03) &0.95 (0.02) &0.99 (0.00) &-\\

\cmidrule(lr){2-11}
&&BNME&0.01 (0.02)&1.00 (0.01)&-&0.96&0.01 (0.01)&1.00 (0.00)&-&1.00 \\
\cmidrule(lr){4-11}
&\textbf{\multirow{2}{*}{100\%}}&LMM&0.58 (0.02)&0.95 (0.01)&-& 0.00&0.25 (0.01)&0.95 (0.01)&-&0.00\\
\cmidrule(lr){4-11}
&&LMEKIN&0.22 (0.02) &0.95 (0.02) &-&0.00 &0.25 (0.01)&0.95 (0.01) &-& 0.02 \\
\cmidrule(lr){4-11}
&&GEMMA&0.22 (0.02)&0.95 (0.01)&-& 0.00&0.25 (0.01)&0.94 (0.01)&-& 0.00\\

\hline
\hline
&&BNME&0.14 (0.03)&0.90 (0.04)&0.92 (0.08)&- &0.09 (0.03)&0.93 (0.05)&0.88 (0.08) &-\\
\cmidrule(lr){4-11}
&\textbf{\multirow{2}{*}{50\%}}&LMM&0.71 (0.23)&0.90 (0.12)&0.62 (0.12)&- &0.32 (0.06)&0.95 (0.05)&0.87 (0.06) &- \\
\cmidrule(lr){4-11}
&&LMEKIN&0.39 (0.11) &0.93 (0.14) &0.70 (0.18) &- &0.23 (0.06) &0.93 (0.04) &0.90 (0.03) &- \\
\cmidrule(lr){4-11}
&&GEMMA&0.39 (0.09)&0.94 (0.03)&0.69 (0.19) &- &0.23 (0.06)&0.95 (0.05)&0.94 (0.04)&- \\

\cmidrule(lr){2-11}
&&BNME&0.21 (0.06)&0.99 (0.01)&0.95 (0.09)&- &0.12 (0.20)&0.99 (0.02)&1.00 (0.00)&- \\
\cmidrule(lr){4-11}
\textbf{\multirow{2}{*}{3}}&\textbf{\multirow{2}{*}{90\%}} &LMM&0.71 (0.23)&0.94 (0.12)&0.73 (0.13)&- &0.25 (0.01)&0.95 (0.01)&0.97 (0.03)&- \\
\cmidrule(lr){4-11}
&&LMEKIN&0.23 (0.10) &0.95 (0.09) &0.85 (0.09) &- &0.21 (0.05) &0.96 (0.02) &0.95 (0.05)&-  \\
\cmidrule(lr){4-11}
&&GEMMA&0.21 (0.06)&0.96 (0.07)&0.85 (0.09)&- &0.21 (0.05)&0.95 (0.01)&0.99 (0.01)&- \\

\cmidrule(lr){2-11}
&&BNME&0.01 (0.01)&1.00 (0.00)&-&1.00&0.02 (0.04)&1.00 (0.02)&-&0.90 \\
\cmidrule(lr){4-11}
&\textbf{\multirow{2}{*}{100\%}}&LMM&0.58 (0.02)&0.94 (0.01)&-& 0.00&0.25 (0.01)&0.94 (0.01)&-&0.00\\
\cmidrule(lr){4-11}
&&LMEKIN&0.22 (0.09)&0.95 (0.03) &-& 0.00&0.21 (0.15) &0.95 (0.04) &-& 0.03 \\
\cmidrule(lr){4-11}
&&GEMMA&0.23 (0.13)&0.95 (0.01)&-&0.01&0.24 (0.01)&0.94 (0.01)&-&0.01 \\

     \bottomrule   \\
     \multicolumn{11}{l}{\small *Phenotypic sensitivity does not exist at a 100\% sparse level with no connection associated with the genotype. } \\
\end{tabular}}
\end{table}

\section{Real data application}\label{sec:data}

\subsection{Imaging genetics data for HCP}\label{sec:hcp}

We implement our model to the imaging genetics data from the Human Connectome Project (HCP).  HCP is a landmark study of healthy subjects that has collected a rich set of imaging, behavioral and genetic data. In the current analyses, we adopt the WU-Minn HCP minimally processed S1200 release that includes over 1,000 healthy young adults aged 22 to 37 years. For each subject, both T1 magnetic resonance imaging (MRI) and diffusion MRI (dMRI) are available, allowing the construction of brain structural connectivity to capture the white matter fiber tracts connecting different brain regions. Specifically, based on the minimally prepossessed dMRI and T1 data from ConnectomeDB, we first generate the whole-brain tractography for each subject via a recent probabilistic tractography pipeline~\citep{data1}. We then perform the anatomical parcellation via Desikan-Killiany (DK) atlas~\citep{data1_3} including 68 cortical surface regions and 19 subcortical regions.  To extract the streamlines linking each pair of ROIs, a series of steps including dilation of each gray matter ROI to incorporate white matter regions, separation of the streamlines connecting several ROIs into parts, and removing obvious outlier streamlines are conducted. Subsequently, the mean fractional anisotropy (FA) value  along streamlines is used to evaluate the strength of structural connections. Eventually, we construct brain structural connectivity for 1,065 subjects. 
 Comprehensive details are available elsewhere on HCP neuroimaging protocols~\citep{van2013wu} and our tractography pipeline~\citep{zhao2023genetic}.

The young adult participants in HCP were also genotyped by Illumina's MultiEthnic Global Array (MEGA) Chip and three specialized neuroimaging chips: Psych, NeuroX, and Immunochip. After standard data quality by excluding subjects with more than 10\% missing SNPs or sex check failure, 1,010 subjects with both genotypes and phenotypes are included in our analyses. For the genetic variants, to mitigate computational cost, we focus on the 1,860 SNPs that were identified in the previous study to highly associate with brain structural network~\citep{zhao2023genetic}. However, unlike the previous analyses that didn't accommodate the sample relatedness, we consider family structure after creating the kinship matrix for 149 pairs of genetically-confirmed monozygotic twins (298 participants), 94 pairs of genetically-confirmed dizygotic twins (188 participants) and their non-twin siblings (524 participants). Besides the proposed BNME, we also implement GEMMA to conduct the genet ic association analyses under our related samples. All the model implementations closely follow the simulation studies, and we account for age, gender, and the top ten genetic principal components in  analyses.

\iffalse

 We apply our network response model to HCP-YA data to investigate the genetic association with brain network phenotype. To construct structural connectivity, we selected 87 ROIs using DK parcellation, which included 68 cortical surfaces regions and 19 subcortical regions. After tracking fibers, identifing connections between ROIs, we summarized the structural connections in each subject using $87 \times 87$ connectivity matrix. As for the genetic variation, a total of 1860 SNPs are considered genetic risk factors. Additionally, we adjusted age, gender, and the first 10 genetic PC scores for the 1860 SNPs. We match the brain connectivity information with kinship matrix in the HCP-YA data and include 1010 subjects in our analysis.
\fi

\subsection{Analysis results}\label{sec:result}
Our goal is to identify risk genetic markers and their associated brain connectivity phenotypic components. Based on the posterior samples of $\eta$, we identify nine risk SNPs as shown in Table~\ref{table:analysis}.
After mapping those SNPs to the genes they belong to, we identify five unique gene variants including THSD7B, LINC01503, LOC105373693, CDH13 and SLC38A8. Among them, THSD7B and CDH13 have been considered to play an essential role in the development of the central nervous system and neural connectivity~\citep{li2022exploring,wang2011meta,polanco2021differential}. Particularly, THSD7B has also been shown to associate with intellectual disability~\citep{lyons2022cluster}; and CDH13 is related to various psychiatric disorders including ADHD and substance abuse~\citep{rivero2013impact,treutlein2011genome}.  To evaluate the neurogenetic processes of the selected genetic variants, we further perform a brain tissue-specific expression quantitative trait loci (eQTL) analysis via the UK Brain Expression Consortium \cite[UKBEC,][]{ramasamy2014genetic}. The consortium generated genotype and exon-specific expression data for 134 neuropathologically healthy subjects under ten different brain tissues, which allows us to evaluate each identified genetic variant on its alteration of tissue-specific and cross-tissue gene expressions within 100kb of the SNP~\citep{zhao2023genetic}. Table~\ref{table:analysis} Column 3 shows the   cross-tissue cis-effect p-value calculated in their BRAINEAC web server, and the regulated genes for each risk SNP. The small p-values of cross-tissue eQTLs reflect the molecular regulation through gene expression over different brain areas, consistent with the genetic association with brain network phenotypes.

\begin{table}[H] \caption{Significant genetic variants and their associated phenotypic structural subnetworks, along with cis-eQTL results obtained from UKBEC brain database.}\label{table:analysis}
\centering
\vspace{0cm}
 \setlength\extrarowheight{1.5pt}
\resizebox{1\textwidth}{!}{
   \begin{tabular}{cccccc} 
\toprule 
&\multicolumn{3}{c}{eQTL} & \multicolumn{2}{c}{Phenotypic subnetworks}\\
\cmidrule(lr){2-4} \cmidrule(lr){5-6}
SNP &Chromosome &p-value & Regulated genes &\# Association & Macroscale systems\\
\cmidrule(lr){1-4} \cmidrule(lr){5-6}
rs2465095	&2	&9.30E-03	&THSD7B &91 &Subcortical, Parietal lobe \\
\hline
rs1918367	&2	&3.50E-02	&GALNT13 &91 &Subcortical, Parietal lobe \\
\hline
rs4725467	&7	&2.20E-02	&GALNTL5 &325 &Subcortical, Temporal lobe \\
\hline
rs10760611	&9	&5.00E-03	&ASB6 &20 &Subcortical \\
\hline
rs4948428	&10	&2.50E-02	&TMEM26 &6 &Subcortical \\
\hline
rs1537969	&13	&5.50E-02	&SGCG &22 &Subcortical, Temporal lobe \\
\hline
rs9928439	&16	&2.50E-02	&SLC38A8 &91 &Subcortical, Temporal lobe \\
\hline
rs6563992	&16	&1.30E-03	& ATP2C2 &15 &Subcortical \\
\hline
rs58090793	&16	&3.30E-03	&ZDHHC7 &3 & Frontal lobe\\
\bottomrule   
\end{tabular}}
\end{table}

\begin{figure}[H]
\centering
  \includegraphics[scale=0.9,clip, trim=0cm 0cm 4cm 0cm, width=1.00\textwidth]{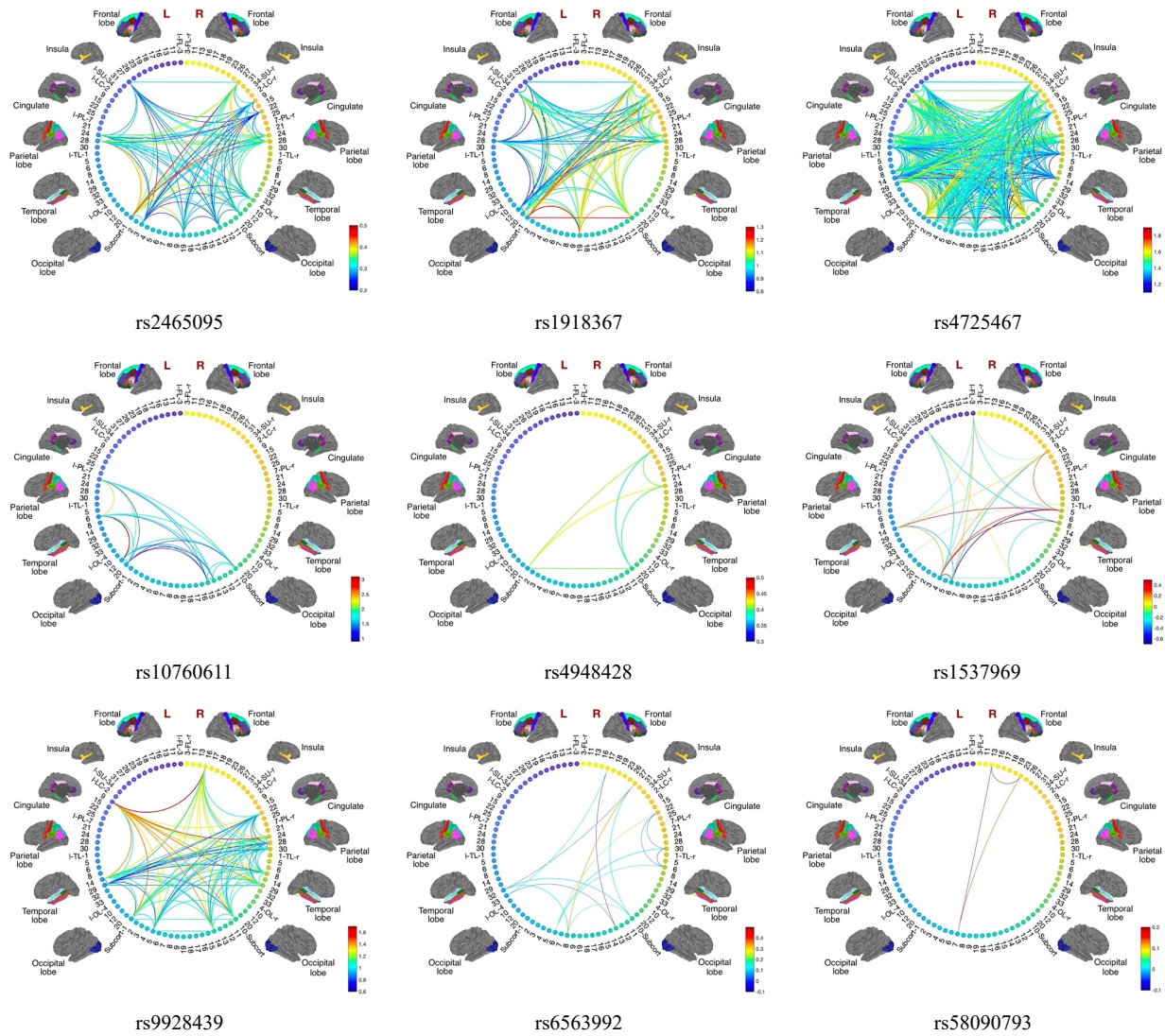}
 \caption{Nine SNPs and their associated brain network.}\label{fig:vis}
\label{cir9}
\end{figure}

We further investigate the associated brain subnetwork phenotypic components for each of the identified genetic signals. Visualization of each genetically associated brain network component is displayed in Figure~\ref{fig:vis},  where the color of connections indicates the effect size of genetic association. Additionally, we summarized the macroscale structures involved in subnetworks for each identified SNP in Table~\ref{table:analysis}. Our analysis revealed that cross-hemispheric connections and inter-subcortical connections accounted for the largest proportion of all the signaling connections. This finding aligns closely with previous literature, which has consistently demonstrated that genetic effects lead to alterations in white matter fiber tracts across brain hemispheres and subcortical structures~\citep{jahanshad2013genome,zhong2021interhemispheric,zhao2021common}. For instance,  \cite{zhong2021interhemispheric} investigated interhemispheric connectivity and highlighted the influence of genetic factors on the integrity and organization of connections between the two hemispheres. Similarly, \cite{hibar2015common} explored the impact of common genetic variations on the subcortical brain  and emphasized the role of genetics in shaping the variations of subcortical structures. 

\begin{figure}[H]
\centering
  \includegraphics[scale=0.2,clip, trim=0cm 5cm 0cm 4cm, width=1\textwidth]{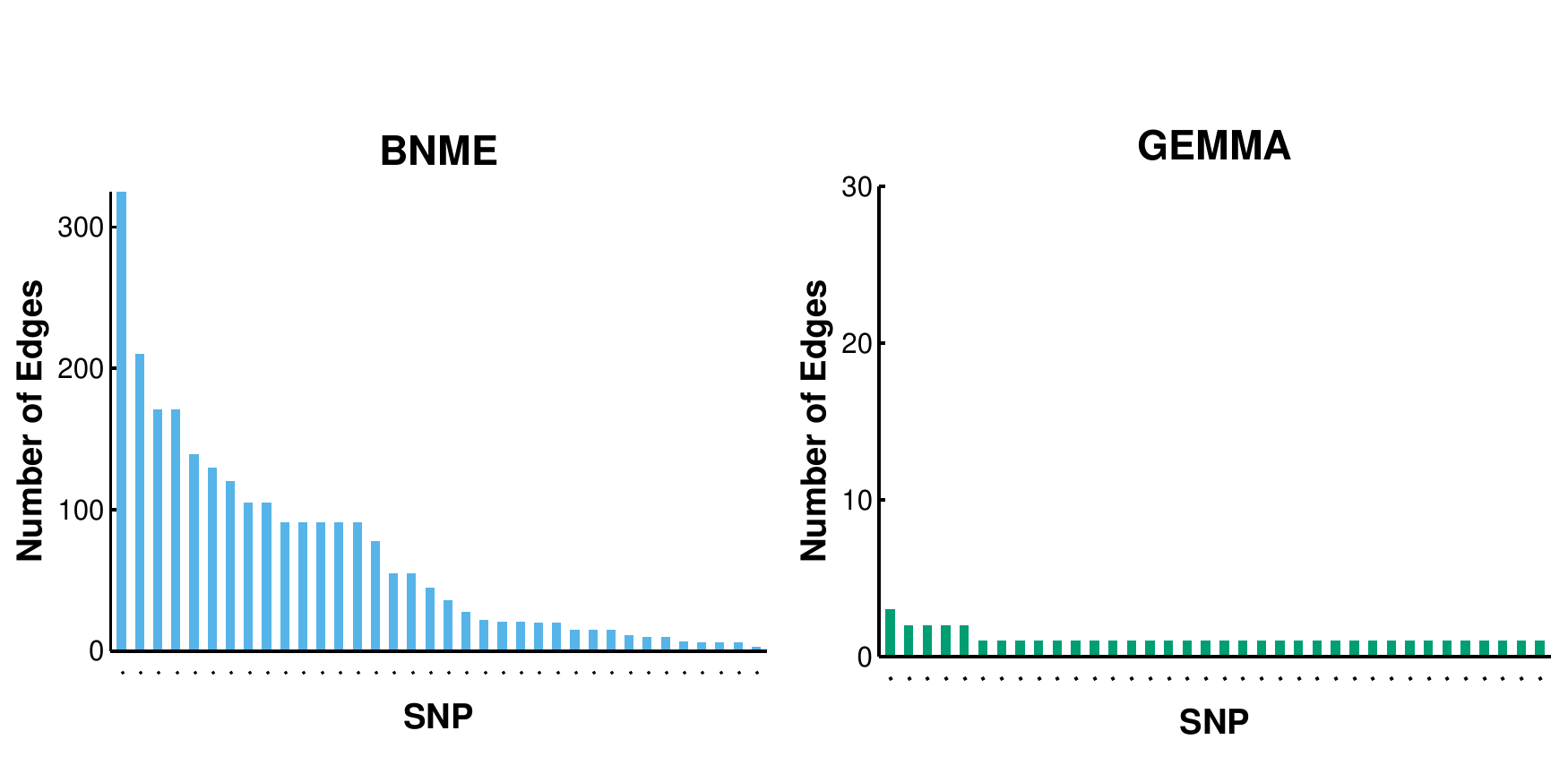}
 \caption{The number of significant edges within the phenotypic network identified by BNME and GEMMA}\label{fig:comp}
\end{figure}

Finally, we conduct a comparison of imaging genetics results between our BNME model and GEMMA. GEMMA identifies a total of 35 SNPs that exhibit significant associations with at least one brain connection. To assess the agreement in the top selected genetic variants between the two approaches, we map the top 35 selected SNPs from each method to their associated cytogenetic bands~\citep{phy_band} and examine the overlap in signals. 
Eventually, there are ten cytogenetic bands that encompass the genetic signals identified by both BNME and GEMMA. This indicates a certain degree of consistency in the genetic signals identified by the two methods, which lends support to the plausibility and reliability of our results.  The detailed results are provided in the supplementary materials. 
\iffalse
are shown in the Table \ref{table:band}. For instance, there are three SNPs-rs4948428, identified by BNME and  rs10509117, rs2448358 identified by GEMMA, locating on long arm of chromosome 10q21.2, which band is associated with childhood acute lymphoblastic leukemia \citep{ch_all}. Additionally, each method identifies series of SNPs on the unique band. For GEMMA, there are multiple SNPs locating on 15q21.3 (rs10518808, rs16976371, rs79506154) and 15q26.1 (rs7168935, rs901966, rs16948292). For BNME, four SNPs (rs9928439, rs12933060, rs6563992, rs58090793) locate on the long arm of chromosome band 16q23.3. 

\begin{table}[H] \caption{cytogenetic bands of SNPs identified by BNME and GEMMA}\label{table:band}
\centering
\vspace{0cm}
 \setlength\extrarowheight{0pt}
\resizebox{0.8\textwidth}{!}{
   \begin{tabular}{lll} 
\toprule 
Band & \multicolumn{2}{c}{SNP} \\
\cmidrule(lr){2-3} 
&\multicolumn{1}{c}{BNME} &\multicolumn{1}{c}{GEMMA}\\
\hline
1q44 & rs6676019 &rs10924849\\
\hline
 2q22.1&rs1991413, rs2465095 &rs10490734 \\
\hline
 2q35&rs6704859, rs13033033 &rs78326416, rs6704859 \\
\hline
3p26.1 &rs728389, rs2671758 &rs6803795, rs73095951\\
\hline
5q14.1 &rs27650 & rs12019306\\
\hline
8q21.13 &rs17300285 &rs16939047 \\
\hline
10q21.2 &rs4948428 &rs2448358, rs10509117 \\
\hline
11p12 &rs770770 &rs57015952, rs10837935\\
\hline
12q24.32 &rs2541988 &rs73430975 \\
\hline
13q33.1 &rs9518867 &rs612938 \\

\bottomrule   
\end{tabular}}
\end{table}
\fi
Furthermore, we also visualize the number of associated brain connections for the top selected SNPs under each method in Figure~\ref{fig:comp}. It is evident that, in contrast to BNME, which provides phenotypic subnetwork architecture specific to each genetic variant, the phenotypic signals identified under GEMMA appear to be extremely sparse and scattered. This result indicates that the majority of the SNPs identified under GEMMA are associated with a single brain connection, raising questions regarding the biological interpretability and meaningfulness of the observed genetic associations. 

\section{Discussion}\label{s:discuss}
In this paper, we present an innovative Bayesian network-response mixed effect model that addresses the challenges of genetic association studies in brain connectivity. Our model is specifically designed to capture the genetic contributions to phenotypic network configurations while accounting for family structures and unknown sample relatedness.  To accommodate the biological architecture in the network phenotype, we consider the genetic variant influences the phenotype via a set of unknown subnetworks, where the targeted phenotypic subnetworks are uncovered through a hierarchical selection procedure. Through posterior inference, we effectively quantify the uncertainty associated with determining a risk genetic variant and its impact on phenotypic network configurations. Extensive simulations demonstrate the superiority of our method over existing alternatives in estimating genetic effects and identifying relevant phenotypic elements with signaling capabilities. By applying the proposed method to the HCP cohort with excessive family structures, we obtain biologically interpretable results that shed light on uncovering the genetic underpinnings of brain structural connectivity.

\iffalse

For future research, an alternative assumption can be applied on the residual term for our model. Currently, we assume the residual term is normally distributed, but it is likely to be more robust if  the residual term is modeled via a Dirichlet process (DP) mixture model \citep{DPMM}. However, DP model has a growing number of unknown parameters since we need an additional set of parameters to construct DP priors. Increasing modeling complexity may expand the risk of suffering intensive computation and poor posterior mixing with a finite sample size given the complex data structure and relatively high-dimensional feature space. 
\fi

In addition to its applicability to brain connectivity genetics studies,  the proposed BNME model provides a  fundamental framework for mixed-effect models involving network- or matrix-variate outcomes. As data collection in epidemiology and social studies becomes more complex, there is a growing need to analyze network-related or matrix-structured outcomes  arising from related samples caused by pedigree or repeated measurements. By extending the random effect tensor, $\mathcal{B}$, to include an additional dimension corresponding to random slopes, along with the associated variance-covariance component, we can effectively capture more intricate sources of variation and address diverse modeling requirements.

Our current model formulation employs a decomposition of the effect matrix into a series of weighted outer products. This design choice aligns well with the biological assumptions inherent in our application and facilitates the interpretation of results. However, in cases where prior knowledge suggests alternative association structures, such as a modular structure, one can easily modify model~\eqref{eq:matrix1} by adopting a different decomposition approach, such as a stochastic block model.  Moreover, our proposed model can be readily extended to perform heritability analyses for network phenotypes.  As a fundamental quantitative genetic analysis, existing heritability analyses only consider scalar- or vector-variate phenotypes. By adapting our model to this future direction, we could contribute to filling this literature gap and provide valuable insights into the heritability of network-related traits.

	\bibliographystyle{asa}
\baselineskip=10pt
\bibliography{biomsample}

\end{document}